\documentclass[a4paper,11pt]{article}

\pdfoutput=1
\usepackage{amsmath}
\usepackage{graphicx}

\usepackage[english]{babel}
\usepackage[utf8]{inputenc}
\usepackage{pdflscape}
\usepackage{enumerate}
\usepackage{amsbsy}
\usepackage{amsmath} 
\usepackage{graphics}
\usepackage{mathrsfs}
\usepackage{wrapfig}
\usepackage{mathtools}

\usepackage{amsfonts}
\usepackage{pstricks}
\usepackage{color}
\usepackage{setspace}

\usepackage{jcappub_ver} 

\newcommand{\bea}{\begin{eqnarray}} \newcommand{\eea}{\end{eqnarray}}
\newcommand{\el}{\nonumber \\}
\newcommand{\re}[1]{(\ref{#1})}

\newcommand{\pat}{\partial}

\renewcommand{\sec}[1]{section \ref{#1}}
\newcommand{\fig}[1]{figure \ref{#1}}

\renewcommand{\a}{\alpha}
\renewcommand{\b}{\beta}
\renewcommand{\c}{\gamma}
\renewcommand{\d}{\delta}

\newcommand{\ha}{\frac{1}{2}}

\newcommand{\rmd}{\mathrm{d}}

\newcommand{\ie}{i.e.\ }

\newcommand{\Mpl}{M_{{}_{\mathrm{Pl}}}}

\newcommand{\half}{\ha}
\newcommand{\f}{\frac}

\newcommand{\pd}{\partial}

\newcommand{\sg }{\sqrt{-g}}

\newcommand{\sq}{\sqrt{-q}}
\usepackage{accents}
\usepackage{tensor}
\newcommand{\TI}{\indices}
\newcommand{\Ut}{\accentset{\sim}{U}}
\newcommand{\Xt}{\accentset{\sim}{X}}
\newcommand{\Ft}{F}
\newcommand{\nt}{\accentset{\sim}{n}}
\newcommand{\rt}{\accentset{\sim}{r}}

\newcommand{\Epst}{\accentset{\sim}{\epsilon}}
\newcommand{\Etat}{\accentset{\sim}{\eta}}

\title{Inflation with $R_{(\a\b)}$ terms in the Palatini formulation}

\author{Jaakko Annala}
\author{and Syksy R\"{a}s\"{a}nen}

\affiliation{University of Helsinki, Department of Physics and Helsinki Institute of Physics \\ P.O. Box 64, FIN-00014 University of Helsinki, Finland}

\emailAdd{jaakko.annala@helsinki.fi}
\emailAdd{syksy.rasanen@iki.fi}

\abstract{We study inflation with the most general non-degenerate gravitational action that depends on the symmetric part of the Ricci tensor coupled to a scalar field in the Palatini formulation of gravity. We use field redefinitions to shift the effect of the Ricci terms from gravity to the scalar field, and apply the result to slow-roll inflation.

As examples, we consider actions quadratic and cubic in the Ricci tensor. In the quadratic case the results are similar to the case $R+\alpha R^2$ that has been studied earlier: the tensor-to-scalar ratio $r$ can be suppressed by an arbitrary amount, while the scalar spectrum is unaffected. In the cubic case, $r$ can be suppressed by at most a factor of $2/9$, and the change in the scalar spectral index $n_s$ can be large.}

\begin{document}

\begin{flushleft}
	\hfill		 HIP-2021-22/TH \\
\end{flushleft}
 
\setcounter{tocdepth}{3}

\setcounter{secnumdepth}{3}

\maketitle

\section{Introduction} \label{sec:intro}

One modification of general relativity common to many theories is the inclusion of higher order terms in the Riemann tensor in the action, whether from ultraviolet completion of the theory or quantum corrections. These terms have a very different effect depending on the formulation of gravity. In the metric formulation, such terms in general lead to higher order equations of motion and thus to the Ostrogradski instability, with notable exceptions, such as theories where the action depends only on the Ricci scalar $R$ \cite{Woodard:2006nt}, Horndeski theories and beyond Horndeski theories \cite{Kobayashi:2019hrl,Gleyzes:2014dya}. (It may also be that the classical instabilities are cured in the quantum theory \cite{Donoghue:2021eto}.) In contrast, in the Palatini formulation of gravity \cite{Einstein:1925, Hehl:1976, Hehl:1978, Hehl:1981}, there is no Ostrogradski instability, because the equations of motion are first order, although there can still be ghost modes \cite{BeltranJimenez:2019acz}. 

The most effective observational probe of high curvature effects available is the primordial universe, in particular cosmic inflation \cite{Starobinsky:1979ty,Starobinsky:1980te, Kazanas:1980tx, Guth:1981, Sato:1981, Mukhanov:1981xt, Linde:1981mu, Albrecht:1982wi, Hawking:1981fz, Chibisov:1982nx, Hawking:1982cz, Guth:1982ec, Starobinsky:1982ee, Sasaki:1986hm, Mukhanov:1988jd}. Conversely, the predictions of particle physics models of inflation have the caveat that they can be changed by modifications of gravity, such as higher order curvature terms. While the curvature during inflation is well below the Planck scale, $\sqrt{R}/\Mpl<0.9\times10^{-4}$ \cite{Akrami:2018odb}, this may be compensated by large coefficients in the action.

Inflation with the gravitational action $R+\a R^2$ was studied in the Palatini formulation in \cite{Enckell:2018hmo}. A Legendre transformation plus a conformal transformation were used to shift the effect of the $R^2$ term completely to the scalar field, so that the theory can be analysed using the standard formalism for single field inflation. To leading order in slow-roll, the scalar spectrum is unchanged, but the tensor amplitude (and spectral index) are suppressed by a term that can be arbitrarily large. In \cite{Masters_thesis} the analysis was extended to the general quadratic case involving only the symmetric part of the Ricci tensor, $R+\a R^2+\b R_{(\a\b)} R^{(\a\b)}$. (In the Palatini formulation, the Ricci tensor is in general not symmetric.) The conformal transformation has to be generalised to a disformal transformation, but the result is the same as in the $R^2$ case with the replacement $\a\to\a+\b/4$.

We generalise the analysis to an arbitrary non-degenerate action that depends on $R_{(\a\b)}$ coupled to a scalar field, but not its derivatives. We include only the symmetric part of the Ricci tensor, as in this case the theory is known to not contain new degrees of freedom, the new terms just change the relation between existing degrees of freedom \cite{Afonso:2017}. 

In \sec{sec:general} we give the action, perform the Legendre and disformal transformations, and solve for the auxiliary fields in the limit of small kinetic term to reduce the gravitational action to the Einstein--Hilbert form and shift the non-standard behaviour to the scalar field kinetic term and potential. In \sec{sec:inflation} we apply this action to slow-roll inflation and derive the inflationary observables. We find that in contrast to the quadratic case, also the scalar spectrum can change. As an example, we consider inflation with quadratic and quartic potentials. We summarise our findings in \sec{sec:conc}.

\section{General non-degenerate $R_{(\a\b)}$ gravity theory} \label{sec:general}

\subsection{Legendre transformation}\label{sec:legendre}

We work in the Palatini formulation of gravity, where the metric $g_{\a\b}$ and the connection are independent degrees of freedom. We consider the most general action where the gravitational sector depends on the connection only via the symmetric part of the Ricci tensor $R_{(\a\b)}$, coupled to a scalar field $\varphi$ (but not its derivatives) that has a standard kinetic term,
\begin{equation} \label{action1}
	S = \int \rmd^4 x \sg  \left[\half F(g_{\a\b},R_{(\a\b)},\varphi) - \half g^{\a\b} X_{\a\b} -  U(\varphi)\right] \ ,
\end{equation}
where $g\equiv\det{g_{\a\b}}$ and $X_{\a\b}\equiv\pat_\a\varphi\pat_\b\varphi$.

First we perform a Legendre transformation to make the action linear in the Ricci tensor \cite{Magnano:1987zz,Koga:1998un,Afonso:2017}
\begin{equation} \label{actionL}
	\begin{aligned}
		S = \int \rmd^4 x \sg& \half \left[ F\left(g_{\a\b},\Sigma_{\a\b},\varphi\right) + \frac{\pd F}{\pd \Sigma_{\a\b}} ( R_{(\a\b)} -\Sigma_{\a\b} ) - g^{\a\b} X_{\a\b} - 2 U(\varphi)\right] \ ,
	\end{aligned}
\end{equation}
where $\Sigma_{\a\b}$ is an auxiliary field. Let us check under which conditions the action \re{actionL} is equivalent to the original action \re{action1}. Variation with respect to $\Sigma_{\a\b}$ gives
\bea\label{var_Sigma}
	\frac{\pd^2 F}{\pd \Sigma_{\a\b}\pd\Sigma_{\c\d}}\left( R_{\c\d} - \Sigma_{\c\d} \right) = 0 \ .
\eea
If the action is linear in $\Sigma_{\a\b}$, the equivalence is trivial. We assume that this is not the case, so $\pat^2 F/(\pat\Sigma_{\a\b}\pd\Sigma_{\c\d})\neq0$. Assuming that $\pat^2 F/(\pat\Sigma_{\a\b}\pd\Sigma_{\c\d})$ is invertible with respect to the indices $\a\b$ (or equivalently $\c\d$), \ie that the theory is non-degenerate, \re{var_Sigma} gives $\Sigma_{\a\b}=R_{\a\b}$. Inserting this solution back into \re{actionL} recovers the action \re{action1}. The invertibility requirement imposes a constraint on the form of $F$.

We now introduce the field redefinition
\bea
	\label{redefs} \sg \frac{\pd F}{\pd \Sigma_{\a\b}} &\equiv& \sq q^{\a\b} \ ,
\eea
where $q\equiv1/\det(q^{\a\b})=\det(q_{\a\b})$, with $q_{\a\b}$ being the inverse of $q^{\a\b}$. (The inverse has to exist for \re{redefs} to be consistent.) With this field redefinition the action becomes 
\bea\label{action_legendre}
S &=& \int \rmd^4x \sq \half \left\{ q^{\a\b}R_{(\a\b)} - q^{\a\b} \Sigma_{\a\b}(g_{\a\b},\varphi) + \frac{\sg}{\sq}\left[ F(g_{\a\b},\Sigma_{\a\b},\varphi)- g^{\a\b} X_{\a\b} - 2 U(\varphi)\right] \right\} \ . \el
\eea
The field $q_{\a\b}$ plays the role of the metric. The auxiliary field $\Sigma_{\a\b}$ is solved in terms of the new field $q_{\a\b}$ (as well as $\varphi$ and $X_{\a\b}$) from the field redefinition \eqref{redefs}. The original metric $g_{\a\b}$ can be solved in terms of the new field from the action \eqref{action_legendre}.

\subsection{Disformal transformation}

In order to solve for $\Sigma_{\a\b}$ and $g_{\a\b}$ we introduce the disformal ansatz
\bea
  \label{metric_ansatz} g_{\a\b} &=& \gamma_1(\varphi, X_q)q_{\a\b}+\gamma_2(\varphi, X_q) X_{\a\b} \\
  \label{sigma_ansatz} \Sigma_{\a\b} &=& C(\varphi, X_q)q_{\a\b} + D(\varphi, X_q) X_{\a\b} \ ,
\eea
where $X_q\equiv q^{\a\b} X_{\a\b}$. The inverse disformal transformation expresses the new metric $q_{\a\b}$ in terms of the old metric $g_{\a\b}$ and $\varphi$,
\begin{equation}\label{disformaltransform}
	q_{\a\b}=\Gamma_1(\varphi, X_g)g_{\a\b}+\Gamma_2(\varphi, X_g)X_{\a\b} \ ,
\end{equation}
where $X_g \equiv g^{\a\b}X_{\a\b}$.

We get the coefficients $\Gamma_i$ in terms of $\gamma_i$ by plugging $q_{\a\b}$ into the expression \eqref{metric_ansatz} for $g_{\a\b}$ and requiring that the result equals $g_{\a\b}$. This gives
\begin{equation}\label{func_rel}
	\Gamma_1(\varphi, X_g) = \frac{1}{\gamma_1\left[\varphi, X_q(\varphi, X_g)\right]} \ , \quad \Gamma_2(\varphi, X_g) = -\frac{\gamma_2\left[\varphi, X_q(\varphi, X_g)\right]}{\gamma_1\left[\varphi, X_q(\varphi, X_g)\right]} \ .
\end{equation}
The inverse of $q_{\a\b}$ is found from the definition $q_{\a\mu}q^{\mu\b}=\delta\TI{^\a_\b}$, which gives
\begin{equation}\label{ansatz_inverse}
	q^{\a\b} = \frac{1}{\Gamma_1}g^{\a\b} - \frac{\Gamma_2}{\Gamma_1(\Gamma_1+\Gamma_2 X_g)} g^{\a\mu}g^{\b\nu}X_{\mu\nu} \ .
\end{equation}
As $(g_{\a\b},\Gamma_i)$ and $(q_{\a\b},\c_i)$ are in a symmetric position, the inverse of the old metric $g_{\a\b}$ in terms of the new metric has the same form
\bea \label{ansatz_inverse_g}
	g^{\a\b} &=& \frac{1}{\gamma_1}q^{\a\b} - \frac{\gamma_2}{\gamma_1(\gamma_1+\gamma_2 X_q)} q^{\a\mu}q^{\b\nu}X_{\mu\nu} \el
	&\equiv& A q^{\a\b} + B q^{\a\mu}q^{\b\nu}X_{\mu\nu} \ ,
\eea
where we have for later convenience introduced the symbols $A$ and $B$. We find $X_q$ as a function of $X_g$ from \eqref{ansatz_inverse},
\begin{equation}\label{kin_qg}
	X_q = \frac{X_g}{\Gamma_1+\Gamma_2 X_g} \ .
\end{equation}
Again, as the old and the new metric are in a symmetric position, we correspondingly have
\begin{equation}\label{Xg}
	X_g = \frac{X_q}{\gamma_1+\gamma_2 X_q} \ .
\end{equation}
We also need the determinant of the old metric $g_{\a\b}$ in terms of the determinant of the new metric. With the help of the matrix determinant lemma, \eqref{metric_ansatz} gives
\begin{equation}\label{detg}
	g = q \gamma_1^3\left( \gamma_1 + \gamma_2 X_q \right).
\end{equation}

The transformation \re{disformaltransform} has to satisfy a set of requirements to ensure that it is a map between two pseudo-Riemannian spaces that are physically equivalent \cite{Minamitsuji:2014waa,Tsujikawa:2014uza,Motohashi:2015pra,Watanabe:2015uqa,Domenech:2015hka,Fumagalli:2016afy,Takahashi:2017zgr,Chiba:2020mte}:
\begin{enumerate}
	\item The disformal transformation is invertible. 
	\item The metric $q_{\a\b}$ has Lorentzian signature.
	\item Causal trajectories remain causal.
\end{enumerate}
The transformation is invertible when the determinant $\left|\frac{\pd q_{\a\b}}{\pd g_{\c\d}}\right|$ of the Jacobian of the transformation \eqref{disformaltransform} is non-zero, which is equivalent to \cite{Zumalacarregui:2013pma}
\begin{equation}\label{Jacobian}
	\Gamma_1\left(\Gamma_1- X_g\frac{\pd \Gamma_1}{\pd X_g}- X_g^2\frac{\pd\Gamma_2}{\pd X_g}\right)\neq 0 \ .
\end{equation}
This also guarantees that the inverse $q^{\a\b}$ given in \re{ansatz_inverse_g} is well defined, i.e. $\Gamma_1 \neq 0$ and $\Gamma_1+X_g \Gamma_2 \neq 0$.

Consider the line element
\begin{equation}\label{line_element}
	\rmd s_q^2 \equiv q_{\a\b} \rmd x^\a \rmd x^\b = \Gamma_1 \rmd s^2_g + \Gamma_2 (\pd_\a \varphi \rmd x^\a)^2 \ ,
\end{equation}
where $\rmd s_g^2\equiv g_{\a\b} \rmd x^\a \rmd x^\b$. Causal trajectories remain causal, \ie $\rmd s^2_g \leq 0$ guarantees $\rmd s^2_q\leq0$, precisely when $\Gamma_1>0$ and $\Gamma_2\leq0$. 

The last condition is that Lorentzian signature is retained. Let us first consider the case when $\pd_\a \varphi$ is timelike. Then we can write the line element \re{line_element} in the ADM decomposition as
\begin{equation}
	\rmd s_q^2 = -N^2(\Gamma_1+X_g \Gamma_2)\rmd t^2 + \Gamma_1 g_{ij}(\rmd x^i + N^i \rmd t)(\rmd x^j + N^j \rmd t) \ ,
\end{equation}
where $N$ is the lapse and $N^i$ is the shift. The Lorentzian signature is preserved when $\Gamma_1 > 0$ and $\Gamma_1+X_g \Gamma_2 > 0$. If $\pd_\a \varphi$ is null, we get the same condition, which reduces to $\Gamma_1>0$ as $X_g=0$. If $\pd_\a \varphi$ is spacelike, we can write the line element as
\begin{equation}
	\rmd s_q^2 = -N^2\Gamma_1 \rmd t^2 + (\Gamma_1 g_{ij} + \Gamma_2 \pd_i \varphi\pd_j \varphi)(\rmd x^i + N^i \rmd t)(\rmd x^j + N^j \rmd t) \ .
\end{equation}
The signature is preserved if $\Gamma_1 > 0$ and $\Gamma_1 g_{ij} + \Gamma_2 \pd_i \varphi\pd_j \varphi$ is positive definite, which means that $(\Gamma_1 g_{ij}+\Gamma_2\pd_i \varphi\pd_j \varphi)v^i v^j >0$ for all non-zero vectors $v^i$ that point along the spatial slice. Assuming $\Gamma_2 \leq 0$ (as required for causality) we have $(\Gamma_1 g_{ij}+\Gamma_2\pd_i \varphi\pd_j \varphi)v^i v^j \geq v^2(\Gamma_1+X_g\Gamma_2)$, so the same conditions as in the timelike and null case, $\Gamma_1 > 0$ and $\Gamma_1+X_g \Gamma_2 > 0$, guarantee that the signature is preserved. In terms of the parameters $\gamma_1,\gamma_2$ of the inverse transformation \eqref{metric_ansatz}, the signature conditions and $\Gamma_2\leq0$ read
\begin{equation}\label{req_phys}
	\gamma_1 > 0 \ ,\quad \gamma_2\geq0\ ,\quad \gamma_1+X_q\gamma_2>0 \ .
\end{equation}
In terms of these parameters, the invertibility condition 1 discussed above is just \eqref{Jacobian} with the substitutions $\Gamma_i \to \gamma_i$ and $X_g \to X_q$. 

\subsection{Action to first order in $X_q$}

Let us now solve for $g_{\a\b}$ and $\Sigma_{\a\b}$ using the disformal ansatz \eqref{metric_ansatz} and \eqref{sigma_ansatz}. We cannot find exact solutions for the ansatz coefficients, but we can solve them approximately in the limit of small $X_q$. The smallness of the kinetic term $X_q=X_q/\Mpl^4$ is a requirement for the classical theory to apply, and the term is further suppressed during slow-roll inflation in the long wavelength limit, which we are interested in. The non-trivial condition for the expansion to be valid is that the factors multiplying powers of $X_q$ (which come from the coefficients of $R_{(\a\b)}$, as we will see in detail in \sec{sec:inflation}) are not too large. We assume this and expand
\bea\label{series_ABs}
    &&A(\varphi, X_q) = A_0(\varphi) + A_1(\varphi)X_q + \mathcal{O}(X_q^2) \el
    &&B(\varphi, X_q) = B_0(\varphi) + B_1(\varphi)X_q + \mathcal{O}(X_q^2) \el
    &&C(\varphi, X_q) = C_0(\varphi) + C_1(\varphi)X_q + \mathcal{O}(X_q^2) \el
    &&D(\varphi, X_q) = D_0(\varphi) + D_1(\varphi)X_q + \mathcal{O}(X_q^2) \ .
\eea
Using \eqref{metric_ansatz} and \eqref{sigma_ansatz} we now expand the action \eqref{action_legendre} to first order in $X_q$. Since $F$ is a scalar, it can be written as a function of terms with $n$ powers of $\Sigma_{\a\b}$ contracted with $g^{\a\b}$, where $n$ can take all non-negative integer values. A term of order $n$ reads, factoring out traces of $\Sigma_{\a\b}$, using \eqref{metric_ansatz}, \eqref{sigma_ansatz} and \eqref{series_ABs}, and expanding to linear order in $X_q$,
\bea
   && (g^{\a\b}\Sigma_{\a\b})^l g^{\a_1\b_1}\dots g^{\a_m\b_m} \Sigma_{\a_1\b_2}\Sigma_{\a_2\b_3}\dots \Sigma_{\a_m\b_1} \el
  &=& \left[(4AC)^l+l(4AC)^{l-1}(AD+BC)X_q\right]\left\{4(AC)^m+mA^{m-1} C^{m-1} ( A D + B C ) X_q\right\} + \mathcal{O}(X_q^2) \el
  &=& 4^{l+1}(A_0C_0)^{n}\left[1+\frac{n(A_0D_0+B_0C_0+4A_0C_1+4A_1C_0)}{4A_0C_0}X_q\right] + \mathcal{O}(X_q^2) \el
  &\equiv& 4^{l+1}(\sigma_0+\sigma_1X_q)^n + \mathcal{O}(X_q^2) \el
  &\equiv& 4^{l+1} \sigma^n \ ,
\eea
where $n=l+m$, $l,m$ are non-negative integers, and we have defined $\sigma_0\equiv A_0C_0$, $\sigma_1\equiv (A_0D_0+B_0C_0+4A_0C_1+4A_1C_0)/4$, and $\sigma\equiv \sigma_0 + \sigma_1 X_q$. So, to first order in $X_q$, $\Ft(g^{\a\b},\Sigma_{\a\b},\varphi)$ is a function of $\sigma$ and $\varphi$, 
\bea\label{F_sigma}
\Ft(g^{\a\b},\Sigma_{\a\b}, \varphi) = \Ft(\sigma,\varphi) + \mathcal{O}(X_q^2) \ .
\eea
Using \eqref{detg}, we can write the determinant of $q_{\a\b}$ as $q=g\left[A^3(A+BX_q)\right]^{-1}$. Applying this relation, \eqref{sigma_ansatz}, \re{ansatz_inverse_g} and \re{F_sigma}, the action \eqref{action_legendre} reads to first order
\bea\label{O1st_action}
S &=& \int \rmd^4x \sq \half \left\{ q^{\a\b}R_{(\a\b)}\right. \el 
&&\left. + \frac{\Ft(\sigma_0, \varphi)-2U}{A_0^2} - \frac{4\sigma_0}{A_0}-\frac{1}{A_0}X_q + \frac{1}{A_0^2}\left[\Ft_{,\sigma}(\sigma_0, \varphi)-4A_0 \right]\sigma_1 X_q \right.\el
&&\left. + \frac{1}{A_0^3}\left[2U - \Ft(\sigma_0, \varphi)+2\sigma_0A_0\right]\left(\frac{1}{2}B_0+2A_1\right) X_q + \mathcal{O}(X_q^2) \right\} \ .
\eea
We can now vary the action with respect to the parameters in the ansatz \re{series_ABs} and obtain algebraic equations from which they can be solved. Varying with respect to $A_0$, the zeroth order result is
\bea \label{Sol_A0}
    A_0 = \frac{\Ft(\sigma_0, \varphi)-2U}{2\sigma_0} \ .
\eea
Varying with respect to $\sigma_0$, the zeroth order result is
\bea \label{eomS_0}
    \Ft_{,\sigma}(\sigma_0, \varphi) = 4 A_0 \ .
\eea
Plugging in \eqref{Sol_A0} and \eqref{eomS_0} back into the action \eqref{O1st_action}, the terms that contain $B_0,A_1$ and $\sigma_1$ cancel, and we obtain the simple result
\bea\label{infl_action_general}
S &=& \int \rmd^4x \sq \left[ \half q^{\a\b}R_{(\a\b)} -\frac{4\sigma_0(U, \varphi)}{\Ft_{,\sigma}(\sigma_0, \varphi)} - \frac{2}{\Ft_{,\sigma}(\sigma_0, \varphi)}X_q + \mathcal{O}(X_q^2) \right] \ .
\eea
Combining \eqref{Sol_A0} and \eqref{eomS_0}, the unknown $\sigma_0 = \sigma_0(U, \varphi)$ is solved from the algebraic equation
\bea\label{sig_eq}
  2 \Ft(\sigma_0, \varphi) - \sigma_0 \Ft_{,\sigma}(\sigma_0, \varphi) = 4 U(\varphi) \ .
\eea
To zeroth order in $X_q$, the requirements \eqref{req_phys} for the metric $q_{\a\b}$ to describe the same spacetime as $g_{\a\b}$ reduce to
\bea\label{dis_req}
A_0>0 \quad \Leftrightarrow \quad \Ft_{,\sigma}(\sigma_0) > 0 \ .
\eea

In the action \re{infl_action_general}, the connection appears only in the Einstein--Hilbert term, so its equation of motion gives the Levi--Civita connection $\mathring\Gamma^\c_{\a\b}$ of the metric $q_{\a\b}$ (up to a projective transformation, which is a symmetry of the theory \cite{Hehl:1978}). As we did not use the connection equation of motion so far, we did not have to make any assumption about its symmetries. Therefore the result is independent of whether we impose zero torsion, zero non-metricity, or neither, as is known in the case when $F$ is quadratic in $R_{(\a\b)}$ \cite{Enckell:2018hmo,Masters_thesis}. The equations of motion of the metric and the scalar field are now the same as for the Einstein--Hilbert action with a minimally coupled scalar field in the metric formulation of gravity. It is not trivial that we obtained a result as simple as \re{infl_action_general} together with \re{sig_eq}. For example, were we to allow a direct coupling between $R_{(\a\b)}$ and $X_{\a\b}$, the function $F$ would (to leading order) depend not only on $\sigma$ and $\varphi$, but also on a function of $\sigma_0$ multiplying $X_q$.

\section{Inflation} \label{sec:inflation}

\subsection{Observables} \label{sec:obs}

Let us now consider the effect on observables in slow-roll inflation. It is convenient to make the kinetic term canonical with the field redefinition
\bea\label{redefXi}
    \frac{\rmd\varphi}{\rmd \chi} = \pm \ha \sqrt{\Ft_{,\sigma}(\sigma_0)} \ .
\eea
In terms of the canonical field, the action \re{infl_action_general} reads
\bea \label{matter_redef}
        S &=& \int \rmd^4x \sq \left[\half q^{\a\b} R_{\a\b} - \Ut -\half q^{\a\b}\pd_\a\chi\pd_\b\chi + \mathcal{O}(\Xt_q^2) \right] \ ,
\eea
where $\Xt_q \equiv q^{\a\b}\pd_\a\chi\pd_\b\chi$, and the effective potential is
\bea\label{U_eff}
  \Ut(\chi) &\equiv& \frac{4\sigma_0}{\Ft_{,\sigma}} \ ,
\eea
where $\sigma_0=\sigma_0\big\{U[\varphi(\chi)], \varphi[\chi]\big\}$. The first two slow-roll parameters are
\bea \label{epsilonandeta}
  \tilde{\epsilon} &\equiv& \half\left(\frac{\Ut_{,\chi}}{\Ut}\right)^2 = \frac{4}{\Ft_{,\sigma}}\left(\frac{U}{\sigma_0}\right)^2\left[1 + \frac{\sigma_0^2\Ft_{,\sigma\varphi}\Ft_{,\sigma\sigma}-2\Ft_{, \varphi}\Ft_{,\sigma}}{4\sqrt{2\epsilon}U(\Ft_{,\sigma}-\sigma_0\Ft_{,\sigma\sigma})}\right]^2 \epsilon \el
    \tilde{\eta} &\equiv& \frac{\Ut_{,\chi\chi}}{\Ut} = \left(\frac{U}{\sigma_0}\right) \eta - \frac{3 \Ft_{,\sigma\sigma} U^2 \sigma_{0,U}}{\Ft_{,\sigma}\sigma_0}  \epsilon + \frac{\sqrt{2\epsilon}U}{8\sigma_0 \Ft_{,\sigma}(\sigma_{0,U})^2}\Bigg\{ 4 (\sigma_{0,U})^2 \sigma_{0,U\varphi}\Ft_{,\sigma}^2 \el 
    && +6\sigma_0(\sigma_{0,U})^3 \Ft_{,\sigma\sigma}\left[ \Ft_{,\sigma\varphi}+\sigma_{0,\varphi}\Ft_{,\sigma\sigma} -4\sigma_0 (\sigma_{0,U})^2 \sigma_{0,U\varphi}\Ft_{,\sigma}\Ft_{,\sigma\sigma} -16\sigma_{0,\varphi}\sigma_{0,UU}\Ft_{,\sigma} \right] \el
    && -(\sigma_{0,U})^3\Ft_{,\sigma}\left( 3\Ft_{,\sigma\varphi}+6\sigma_{0,\varphi}\Ft_{,\sigma\sigma}+4\sigma_0\Ft_{,\sigma\sigma\varphi}  \right)  \Bigg\} \el
    &&+\frac{1}{8\sigma_0}\Bigg\{ \frac{8\sigma_{0,\varphi\varphi}}{\sigma_{0,U}}+\frac{3\sigma_0\Ft_{,\sigma\varphi}^2}{\Ft_{,\sigma}}-2\sigma_0\Ft_{,\sigma\varphi\varphi} - (\sigma_{0,\varphi})^2\left[ 3\Ft_{,\sigma\sigma}-\frac{3\sigma_0\Ft_{,\sigma\sigma}^2}{\Ft_{,\sigma}}+8\frac{\sigma_{0,UU}}{(\sigma_{0,U})^3} \right] \el
    &&-\sigma_{0,\varphi}\Ft_{,\sigma\varphi}\left( 3-\frac{6\sigma_0\Ft_{,\sigma\sigma}}{\Ft_{,\sigma}}\right)-4\sigma_{0,\varphi}\sigma_0\Ft_{,\sigma\sigma\varphi} \Bigg\} \ ,
\eea
where $\epsilon\equiv\half \left(\f{U_{, \varphi}}{U}\right)^2$ and $\eta\equiv\frac{U_{, \varphi\varphi}}{U}$ are the first two slow-roll parameters in the Einstein--Hilbert case $\Ft=g^{\a\b}R_{(\a\b)}$. It is straightforward to compute all the slow-roll parameter by taking derivatives of the potential. However, when we allow for non-minimal coupling of the scalar $\varphi$ to the Ricci tensor, these expressions are rather cumbersome. In the case of minimal coupling, we have $\Ft=F(\sigma)$ and $\sigma_0=\sigma_0(U)$, and the first two slow-roll parameters \re{epsilonandeta} reduce to
\bea\label{t_eps}
    \Epst &=& \frac{4}{\Ft_{,\sigma}}\left(\frac{U}{\sigma_0}\right)^2 \epsilon \el
    \Etat &=& \left(\frac{U}{\sigma_0}\right) \eta - \frac{3 \Ft_{,\sigma\sigma} U^2 \sigma_{0,U}}{\Ft_{,\sigma}\sigma_0}  \epsilon \ .
\eea
The power spectrum and tilt of the scalar perturbations are, respectively,
\bea\label{t_As}
    \tilde{\mathcal{P}}_\mathcal{R} &=& \left(\frac{\sigma_0}{U}\right)^3 \mathcal{P}_\mathcal{R} \el
    \nt_s-1 &=& \frac{U}{\sigma_0} (n_s-1) + 6 \frac{U}{\sigma_0} \left( 1-\frac{4+\sigma_0\Ft_{,\sigma\sigma} \sigma_{0,U}}{\Ft_{,\sigma}} \frac{U}{\sigma_0} \right) \epsilon \ ,
\eea
where $\mathcal{P}_\mathcal{R} = \frac{U}{24\pi^2\epsilon}$ and $n_s-1=-6\epsilon+2\eta$. The power spectrum and tilt of the tensor perturbations are, respectively,
\bea\label{t_At_nt}
    \tilde{\mathcal{P}}_T &=& \frac{4}{\Ft_{,\sigma}} \frac{\sigma_0}{U} \mathcal{P}_T \el
    \nt_t &=& \frac{4}{\Ft_{,\sigma}}\left(\frac{U}{\sigma_0}\right)^2 n_t \ ,
\eea
where $\mathcal{P}_T = \frac{2U}{3\pi^2}$ and $n_t = -2\epsilon$. The tensor-to-scalar ratio becomes
\bea\label{t_r}
    \rt = \frac{4}{\Ft_{,\sigma}}\left(\frac{U}{\sigma_0}\right)^2 r \ ,
\eea
where $r=16\epsilon$. Finally, the number of e-folds in the slow-roll approximation is
\bea\label{e_folds}
	N_{*} = \int_{\chi_{end}}^{\chi_{*}}\frac{\rmd\chi}{\sqrt{2\Epst}} = \int_{\varphi_{end}}^{\varphi_{*}}\frac{\rmd \varphi}{\sqrt{2\epsilon}} \frac{\sigma_0}{U} \ .
\eea

We have assumed that inflationary observables are invariant under the transformations we have used to go from a higher order curvature action with $g_{\a\b}$ to ordinary gravity with $q_{\a\b}$. If a field transformation such as the disformal transformation is invertible there is a one-to-one mapping between the transformed and original theories, and they are physically equivalent \cite{Fumagalli:2016afy, Takahashi:2017zgr}. This does not necessarily mean that observables such as inflationary power spectra are invariant, but this has been shown to be the case for disformal transformations at least in Horndeski theory \cite{Minamitsuji:2014waa,Tsujikawa:2014uza,Domenech:2015hka,Motohashi:2015pra,Watanabe:2015uqa} (see \cite{Chiba:2020mte} for other observables). For conformal transformations the equivalence between different frames has been shown to all orders in perturbation theory \cite{Catena:2006bd,Gong:2011qe,White:2012ya,Kamenshchik:2014waa,Kubota:2020ehu}. For $f(R)$ theory in the metric formulation, the different field coordinates have been shown to be on-shell equivalent at one-loop order \cite{Ruf:2017xon,Ohta:2017trn}. Similar studies should be done for the case we consider here, including the Legendre transformation, to check whether the observables indeed remain invariant.

To see concretely how the observables change, we have to specify the form of $\Ft$ and solve $\sigma_0$ from \eqref{sig_eq}, keeping in mind that the requirement \eqref{dis_req} has to be satisfied, and that $\Ft$ cannot be degenerate as discussed in \sec{sec:legendre}.

\subsection{Quadratic action} \label{sec:quadratic}

Let us review the quadratic case, which was first covered in \cite{Masters_thesis,Enckell:2018hmo} and applied in \cite{Gialamas:2021enw}:
\bea \label{F_2nd}
	F(g_{\a\b},R_{(\a\b)}, \varphi) &=& g^{\a\b}R_{(\a\b)} +  [ \alpha g^{\a\b} g^{\c\d} + \beta g^{\a\c}g^{\b\d} ] R_{(\a\b)}R_{(\c\d)} \ .
\eea
A function $f(\varphi)$ multiplying the Ricci scalar $g^{\a\b} R_{(\a\b)}$ could be removed with the conformal transformation $g_{\a\b}\rightarrow f(\varphi)^{-1}g_{\a\b}$, leading to a redefinition of $\varphi$ and $U$. (The terms quadratic in $R_{(\a\b)}$ are invariant under the conformal transformation.) So leaving it out does not involve loss of generality. We assume that $\a$ and $\b$ are constants. Applying \eqref{metric_ansatz} and \eqref{sigma_ansatz}, to first order in $X_q$ the function \eqref{F_2nd} becomes
\bea\label{F_2nd_O1}
    \Ft(\sigma) = 4\sigma + 4(4\alpha+\beta) \sigma^2 \ .
\eea
As $\Ft_{,\sigma}=4+8(4\alpha+\beta) \sigma$, the conditions $\sigma_0>0$ and $4\alpha+\beta>0$ are sufficient for the requirement \eqref{dis_req} to be satisfied. If $4\alpha+\beta=0$, only the traceless part of $R_{(\a\b)}$ appears, and $\pat F/(\pat\Sigma_{\a\b}\pat\Sigma_{\c\d})$ is degenerate. Therefore our Legendre transformation does not apply. This case has been analysed using the original form of the action \cite{Masters_thesis}. The result turns out to be same as simply taking $4\alpha+\beta=0$ in the present calculation, \ie there is no change at leading order in $X_q$.

The equation \eqref{sig_eq} that gives $\sigma_0$ as a function of $U$ is now linear, and gives the simple solution $\sigma_0 = U$. The effective potential reads
\bea\label{Ueff_2nd}
  \Ut(\chi) &\equiv& \frac{U[\varphi(\chi)]}{1+2(4\alpha+\beta)U[\varphi(\chi)]} \ .
\eea
Plugging $\sigma_0 = U$ into \eqref{t_eps}--\eqref{e_folds}, we get the first slow-roll parameter  \cite{Masters_thesis}
\bea
  \tilde{\epsilon} &\equiv& \half\left(\frac{\Ut_{,\chi}}{\Ut}\right)^2 = \frac{\epsilon}{1+2(4\alpha+\beta)U} \ .
\eea
As $\tilde\epsilon$ is suppressed by the same factor as $\tilde U$, we have $U/\epsilon=\tilde U/\tilde\epsilon$. So the scalar power spectrum, and all slow-roll parameters derived from it (as well as the number of e-folds \re{e_folds}) are unaffected by the $\a R^2$ and $\b R^{(\a\b)} R_{(\a\b)}$ terms to leading order in slow-roll. Therefore the scalar spectral index and its derivatives are unchanged. However, because the potential is multiplied by $[1+(8\alpha+2\beta)U]^{-1}$, both $n_t$ and $r$ in \re{t_At_nt} and \re{t_r} are suppressed by this factor.

\subsection{Cubic action}\label{sec:cubic}

Let us consider an action that contains terms linear and cubic in $R_{(\a\b)}$:
\bea\label{cubic_F}
    \Ft(g^{\a\b},R_{\a\b}, \varphi) &=& g^{\a\b}R_{\a\b} + \kappa_1 (g^{\a\b}R_{\a\b})^3 + \kappa_2 g^{\mu\nu}R_{\mu\nu} g^{\a\c}g^{\b\d}R_{\a\b}R_{\c\d} \el
    && + \kappa_3 g^{\a\c}g^{\b\mu}g^{\d\nu}R_{\a\b}R_{\c\d}R_{\mu\nu} \ .
\eea
A non-minimal coupling $f(\varphi)$ of the Ricci scalar could again be removed by the conformal transformation $g_{\a\b}\rightarrow f(\varphi)^{-1}g_{\a\b}$, but now the non-minimal coupling would appear in the cubic term, $\kappa_i\to f(\varphi) \kappa_i$. We assume that $\kappa_i$ are constant. Applying \eqref{metric_ansatz} and \eqref{sigma_ansatz}, we have to first order in $X_q$
\bea
    \Ft(\sigma) = 4\sigma + 4 \kappa \sigma^3 \ ,
\eea
where $\kappa \equiv 16\kappa_1+4\kappa_2+\kappa_3$. As $\Ft_{,\sigma} = 4+12\kappa \sigma^2$, the condition $\kappa>0$ is sufficient for the requirement \eqref{dis_req} to be satisfied. From \eqref{sig_eq} we get a cubic equation for $\sigma_0$
\bea\label{sig_eq_3rd}
    \kappa\sigma_0^3-\sigma_0+U=0 \ .
\eea
There are three different branches of solutions depending on the sign of $\kappa$ and $4/\kappa-27U^2$. (See \cite{BeltranJimenez:2020guo} for discussion of branching solutions in Palatini gravity.) In two of the branches either the limit $\kappa\to 0$ does not exist or the effective potential is not bounded from below. We concentrate on the third branch, where $\kappa>0$ and $4/\kappa-27U^2>0$. This implies that $U$ has to be bounded between $-\tfrac{2}{3\sqrt{3\kappa}}<U<\tfrac{2}{3\sqrt{3\kappa}}$. In this case \eqref{sig_eq_3rd} has $3$ real solutions that can be written as
\bea
    \left\{\sigma_0(U)\right\}_n = \frac{2}{\sqrt{3\kappa}}\cos\left[\frac{1}{3}\arccos\left(-\frac{3}{2}\sqrt{3\kappa}U\right)-\frac{2\pi n}{3}\right] \ ,
\eea
where $n=0,1,2$. For $n=0,2$ the limit $\kappa\to 0^+$ does not exist. For $n=1$ we recover $\sigma_0(U)\to U$ in the limit $\kappa\to 0^+$. In this case the effective potential reads
\bea
    \Ut = \frac{2\sqrt{3/\kappa}\sin\left[\frac{1}{3}\arcsin\left(\frac{3}{2}\sqrt{3\kappa}U\right)\right]}{9-6\cos\left[\frac{1}{3}\arccos\left(1-\frac{27}{2}\kappa U^2\right)\right]} \ ,
\eea
which is bounded from both below and above. In the limit of small $\sqrt{\kappa} U$ the leading correction to the effective potential is $\Ut = U - 2\kappa U^3 + \mathcal{O}(\kappa^2 U^5)$. However, the correction need not be small. While we have used an expansion in $X_q$, we have not made any approximation with regard to the value of $\kappa$.

\begin{figure}[t!]\hspace{-0.5cm}
    \includegraphics[scale=0.45]{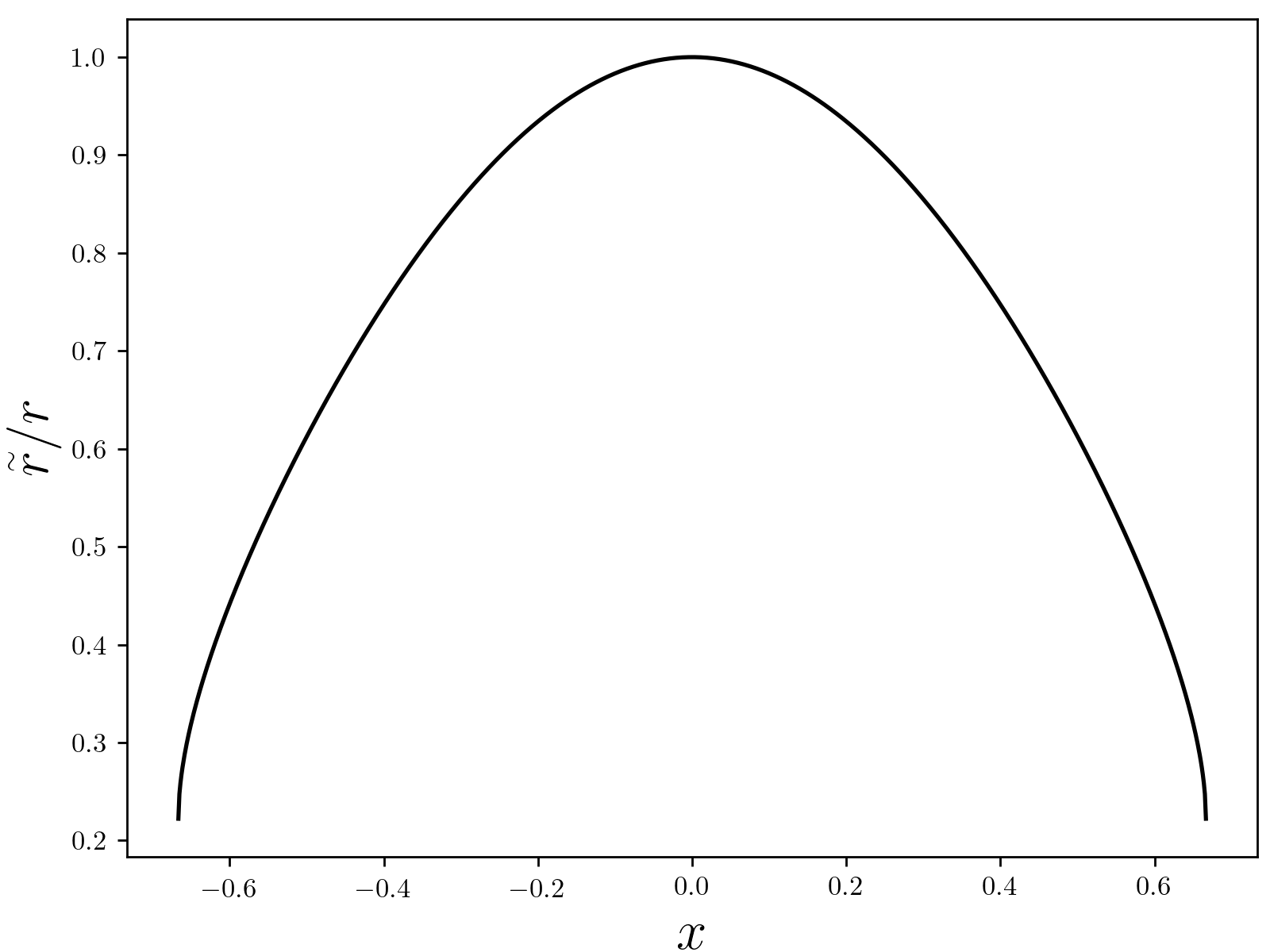}
    \hspace{0.2cm}
    \includegraphics[scale=0.45]{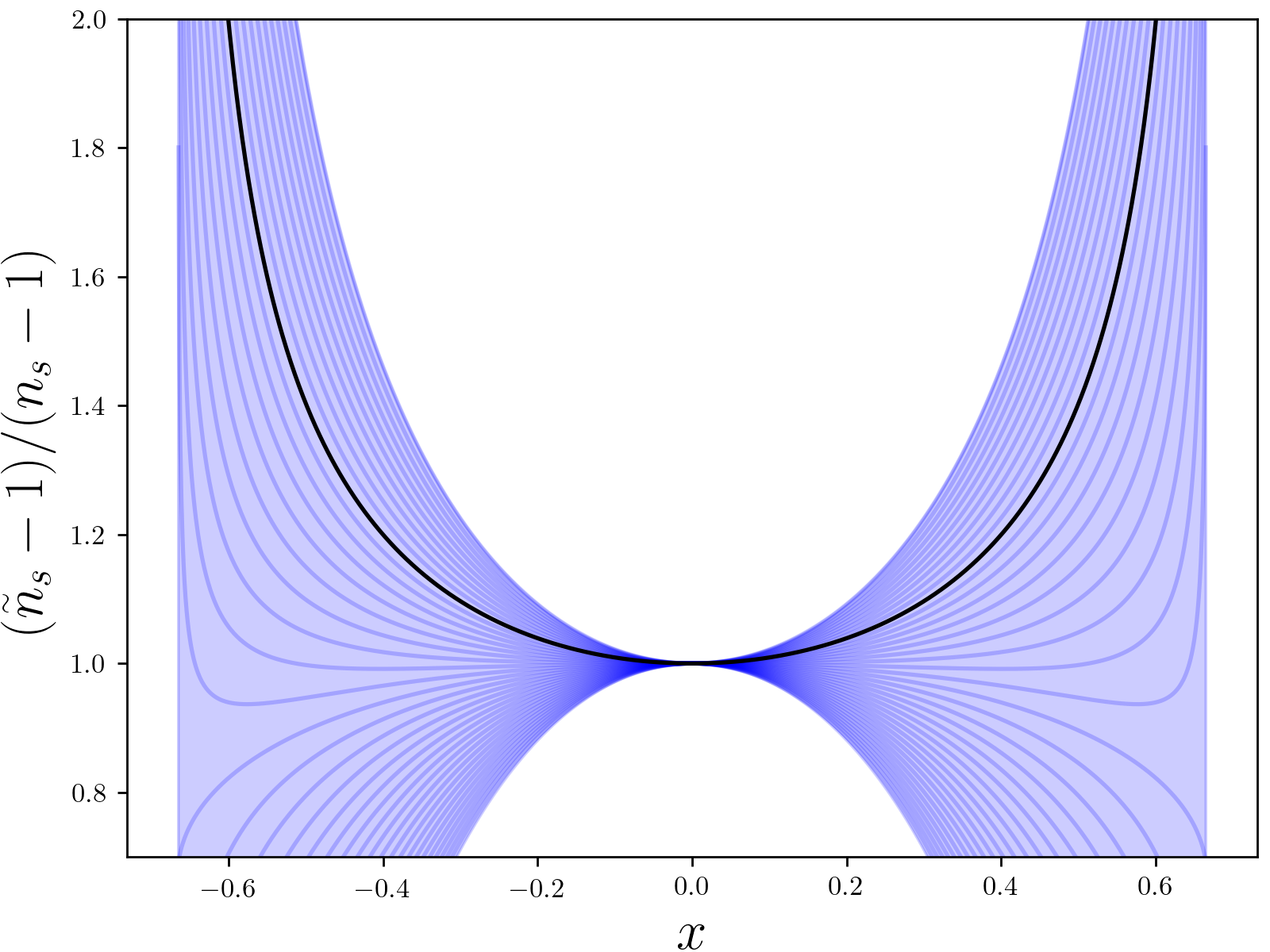}
    \caption{Left: the change to the tensor-to-scalar ratio due to the cubic terms. Right: the change of the scalar spectral index. The blue region shows the range $-10<r/(n_s-1)<10$. The black line corresponds to $r/(n_s-1)=-5$, which approximately corresponds to the potentials $\frac{\lambda}{4}\varphi^2$ and $\frac{m^2}{2}\varphi^4$.}
    \label{fig:nsr}
\end{figure}

We have now solved $\sigma_0$ in terms of $\varphi$. To see how the observables change we just have to plug the solution into the results \eqref{t_As}--\eqref{t_r}. The changes depend on $x\equiv\sqrt{3\kappa}U$, which has the range $-\frac{2}{3}<x<\frac{2}{3}$. For the scalar spectrum we get
\bea\label{scalar_change}
    \tilde{\mathcal{P}}_\mathcal{R} &=& \frac{8}{x^3}\sin^3\left[ \frac{1}{3}\arcsin\left(\frac{3}{2}x\right) \right] \mathcal{P}_\mathcal{R}  \el
   \tilde n_s-1&=& \frac{x}{2\sin\left[\frac{1}{3}\arcsin\left(\frac{3}{2}x\right)\right]} (n_s-1) \\
    &&- \frac{3x^2+12x\cos\left[\frac{1}{3}\arccos\left(-\frac{3}{2}x\right)+\frac{\pi}{3}\right]+6x^2\left(4-9x^2\right)^{-1/2}\sin\left[\frac{1}{3}\arccos\left(-\frac{3}{2}x\right)+\frac{\pi}{3}\right]}{4\sin^4\left[\frac{1}{3}\arcsin\left(\frac{3}{2}x\right)\right]+\sin^2\left[\frac{1}{3}\arcsin\left(\frac{3}{2}x\right)\right]} \epsilon \ . \nonumber
\eea
For the tensor spectrum we have
\bea\label{tensor_change}
  \tilde{\mathcal{P}}_T &=& \frac{2\sin\left[\frac{1}{3}\arcsin\left(\frac{3}{2}x\right)\right]}{x+4x\sin^2\left[\frac{1}{3}\arcsin\left(\frac{3}{2}x\right)\right]} \mathcal{P}_T \el
  \tilde r &=& \frac{x^2}{4\sin^2\left[\frac{1}{3}\arcsin\left(\frac{3}{2}x\right)\right] + 16\sin^4\left[\frac{1}{3}\arcsin\left(\frac{3}{2}x\right)\right]} r \ .
\eea
As in the quadratic case, $r$ is suppressed, but there are two important differences. First, the maximal suppression is $\tfrac{2}{9}$, achieved in the limit $x\to\frac{2}{3}$. Second, the absolute value of $n_s-1$ is boosted without limit when $x\to\frac{2}{3}$. The changes to $r$ and $n_s-1$ are shown as a function of $x$ in \fig{fig:nsr}. 

\begin{figure}[t!]
    \center
    \includegraphics[scale=0.55]{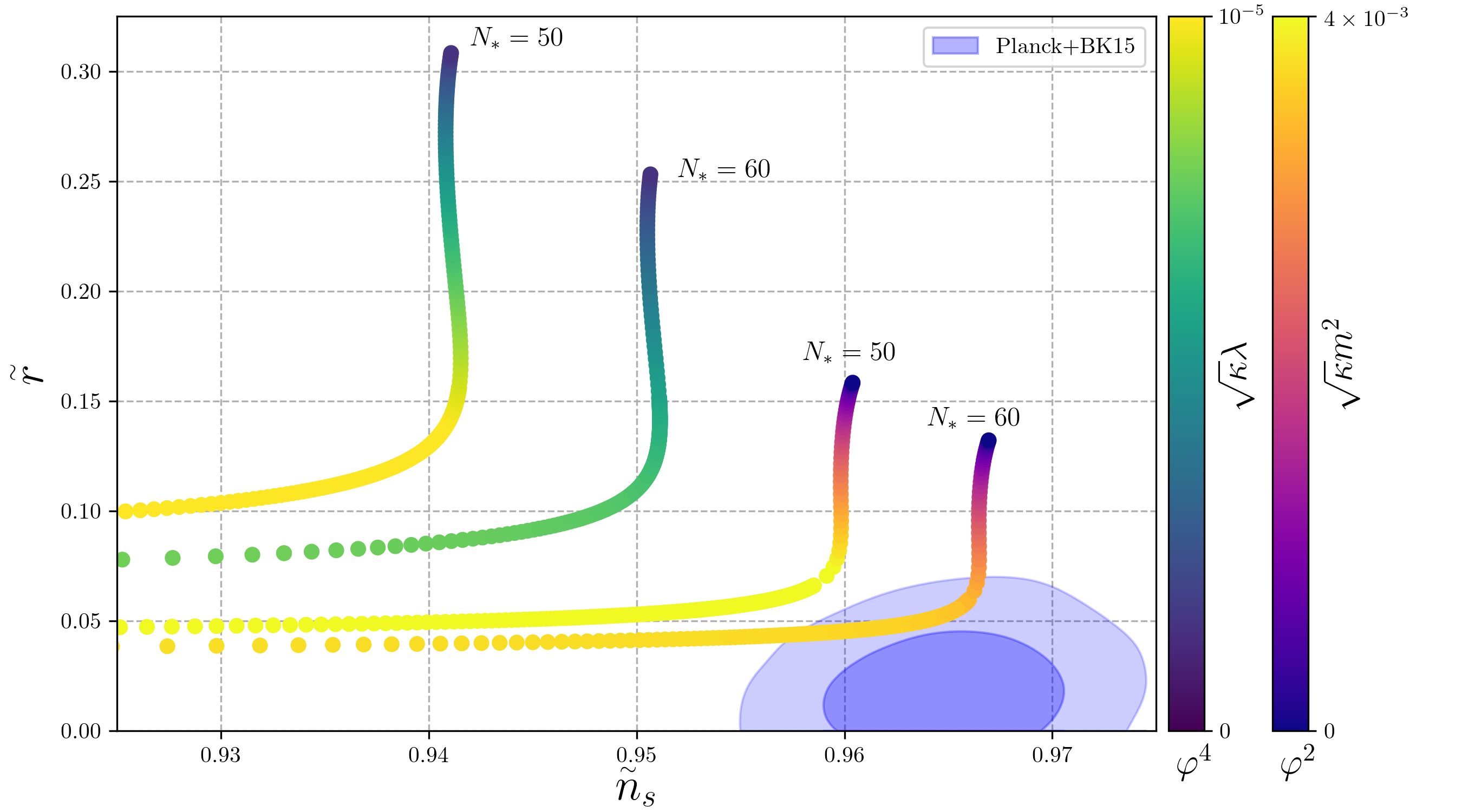}
    \caption{Spectral tilt and tensor-to-scalar ratio. The two curves on the left are for the potential $U=\frac{\lambda}{4}\varphi^4$, with $N_*=50,60$ and different values of $\sqrt{\kappa}\lambda$. The two curves on the right are correspondingly for the potential $U=\frac{m^2}{2}\varphi^2$. The bottom right areas corresponds to the 68\% and 95\% confidence limit regions from Planck and BICEP2/Keck data \cite{Akrami:2018odb}.}
    \label{fig:cubic_nsr}
\end{figure}

For illustration, we consider the potentials $U = \frac{\lambda}{4}\varphi^4$ and $U = \frac{m^2}{2} \varphi^2$. Although $U$ has to satisfy the condition $-\tfrac{2}{3\sqrt{3\kappa}}<U<\tfrac{2}{3\sqrt{3\kappa}}$, we can consider these potentials as approximations valid only below some field amplitude. We numerically solve for the relation between the field value and number of e-folds $N_*$ at the pivot scale $k_*=0.002$ Mpc$^{-1}$. We plot the results for $n_s$ and $r$ for the values $N_*=50,60$ and different values of $\sqrt{\kappa}\lambda$ in \fig{fig:cubic_nsr}.

For the quartic potential, large suppression of $r$ makes $n_s$ even more red and thus more discrepant with observations. In contrast, for the quadratic potential, for which $n_s$ agrees with observations without the cubic Ricci terms and only $r$ needs to be suppressed, the model can be made to agree with the observations for $N_*=60$, with $r$ pushed just below the current upper bound from cosmic microwave background observations.

\section{Conclusions} \label{sec:conc}

In the Palatini formulation of gravity, we have considered the most general non-degenerate action where the connection enters only via the symmetric part of the Ricci tensor, coupled to a single scalar field with a canonical kinetic term. With a Legendre and disformal transformation, we have shifted the effect of Ricci terms from the gravity sector to the scalar field, in the limit when the kinetic term is small. We have further derived the change in inflationary observables.

The case quadratic in the Ricci scalar was derived in \cite{Enckell:2018hmo}, and the general quadratic case was worked out in \cite{Masters_thesis}. Then only the tensor spectrum is modified, the scalar spectrum is unchanged (to leading order in slow-roll). We find that in general also the scalar spectrum changes. As a concrete example, we derive the effective potential in the cubic case, and show that it cannot reconcile the inflationary potential $\frac{\lambda}{4}\varphi^4$ with observations, as the spectral index that is already too red becomes even redder. The adjustment can bring the predictions of the theory with the potential $\ha m^2\varphi^2$ into agreement with observations, just as in the quadratic case. Other forms of the gravitational action could be considered to adjust inflationary predictions as desired. Our results have the caveat that it remains to be shown that the inflationary observables remain invariant under the field transformations, as we have assumed.

Our analysis does not capture possible effects of the curvature terms when the slow-roll approximation is violated. If the scalar field potential alone does not support slow-roll inflation, and slow-roll is only possible when assisted by the higher order curvature terms, our results do not necessarily apply. Also, during preheating the field can roll rapidly. Our calculation could be extended to include multiple scalar fields as well as direct coupling between the Ricci tensor and the scalar field kinetic term, although the resulting action would be more complicated.

As the higher order curvature terms can drastically modify the effective potential for the scalar field, they could change conclusions about the apparent violation of tree-level unitarity in Higgs inflation \cite{Bezrukov:2007, Barbon:2009, Burgess:2009, Burgess:2010zq, Lerner:2009na, Lerner:2010mq, Hertzberg:2010, Bauer:2010, Bezrukov:2010, Bezrukov:2011a, Calmet:2013, Weenink:2010, Lerner:2011it, Prokopec:2012, Xianyu:2013, Prokopec:2014, Ren:2014, Escriva:2016cwl, Fumagalli:2017cdo, Gorbunov:2018llf, Ema:2019, McDonald:2020, Shaposhnikov:2020fdv, Enckell:2020lvn} (anyway different in the Palatini formulation than in the metric formulation \cite{Bauer:2008, Bauer:2010, McDonald:2020, Shaposhnikov:2020fdv, Enckell:2020lvn}), which are sensitive to higher-dimensional operators \cite{Hamada:2020kuy}.

\bibliographystyle{JHEP}
\bibliography{ricci}

\end{document}